\documentclass[aps,pra,twocolumn,showpacs]{revtex4}
\usepackage{bm}
\usepackage{amsmath}
\usepackage{latexsym}
\usepackage{dcolumn}                 % Aligning table columns
\newcommand{\as}[1]{\renewcommand{\arraystretch}{#1}}
\newcommand*{\centt}[1]{\multicolumn{1}{c}{#1}}

\newcolumntype{w}[1]{D{.}{.}{#1}}

\begin{document}
\preprint{Version 2.0}

\title{Nuclear structure corrections in muonic deuterium}

\author{Krzysztof Pachucki}
%\email[]{krp@fuw.edu.pl} \homepage[]{www.fuw.edu.pl/~krp}

\affiliation{Faculty of Physics,
             University of Warsaw,
             Ho\.{z}a 69, 00-681 Warsaw, Poland}

\date{\today}

\begin{abstract}
The muonic hydrogen experiment measuring the $2P-2S$ transition energy 
[R. Pohl {\em et al.}, Nature {\bf 466}, 213 (2010)] is 
significantly discrepant with theoretical predictions 
based on quantum electrodynamics.
A possible approach to resolve this conundrum is 
to compare experimental values with theoretical predictions in
another system, muonic deuterium $\mu$D. The only correction which might be 
questioned in $\mu$D is that due to the deuteron polarizability. 
We investigate this effect
in detail and observe cancellation with the elastic 
contribution. The total value obtained for the deuteron structure
correction in the $2P-2S$ transition is 1.680(16) meV.
\end{abstract}

 \maketitle

The determination of electromagnetic properties of nuclei from precise 
atomic spectroscopy has become possible due to significant progress 
in atomic structure theory, which in turn is based on 
quantum electrodynamics (QED). The proton charge radius as obtained
from the Lamb shift in hydrogen \cite{NIST} is more accurate than any
determination using electron scattering, including the most recent ones \cite{rp_mainz, rp_jlab},
and in agreement with them.
The determination of the deuteron charge radius from the 
measurement of the hydrogen-deuterium isotope shift 
in the 2S-1S transition frequency \cite{H-Da, H-Db},
apart from being the most accurate, has stimulated reanalysis 
of the electron scattering data. At present the atomic isotope shift
determinations of charge radii for helium, lithium and beryllium atoms 
are not only the most accurate ones, but also the only ones available for 
short lived isotopes \cite{be11}.

It was a great surprise that the proton charge radius $r_p$ determined  
from the muonic hydrogen Lamb shift \cite{pohl} gave a result in conflict with the
value determined
from electronic hydrogen. The $5\,\sigma$ discrepancy in $r_p$ is 
the first indication that our knowledge of interactions in
these simple atomic systems is not complete. While we will not pursue
the possible explanations of the proton charge radius discrepancy,
we point attention to the critical test which can be performed
with muonic deuterium. The electronic H-D isotope shift gives a
very accurate difference for the deuteron and proton charge radii.       
If the results for the difference between $\mu$D and $\mu$H are consistent
with the electronic H-D, this would mean that there is an extra 
muon-proton interaction, which cancels out in the $\mu$H-$\mu$D difference.
In order to draw these conclusions, all other effects contributing to
the isotope shift have to be analyzed. The only correction which goes beyond 
the standard QED treatment is that due to the nuclear polarizability, and is
in general due to the nuclear structure. 
It is the purpose of this work to study these effects in muonic deuterium.

Before this, however, we note that there is no unique definition of the charge 
radius for all nuclei. It depends on the nuclear spin,
and in particular, there is no established definition for spin 1 nuclei,
such as the deuteron. The mean square charge radius $\langle R^2 \rangle$
for an arbitrary spin $I$ particle is defined through the effective interaction 
with the electromagnetic field
\begin{eqnarray}
\delta H &=& e\,A^0 - e\,\vec d\cdot\vec E -
e\,\biggl(\frac{\langle R^2 \rangle}{6}+\frac{\delta_I}{M^2}\biggr)\,
\vec\nabla\cdot\vec E
\nonumber \\ &&
-\frac{e}{2}\,Q\,(I^i\,I^j)^{(2)}\,\nabla^j E^i-\vec\mu\cdot\vec B
\label{01}
\end{eqnarray}
where $\mu$ and $Q$ are the magnetic dipole and the electric quadrupole moments.
For a scalar particle $\delta_0 = 0$, and for a half-spin particle $\delta_{1/2} = 1/8$.
For a vector and higher spin particle we proceed as follows. 
The most general Lagrangian for particle with spin $I=1$, which includes
terms linear in the electromagnetic field strength $F$ is \cite{spin1}
\begin{eqnarray}
{\cal L} &=& -\frac{1}{2}\,u^{*\mu\nu}\,u_{\mu\nu} + m^2\,u^{*\mu}\,u_{\mu}
\nonumber \\&&
             +\frac{i\,e}{2}\,(g-1)\,(u^*_\mu\,u_\nu-u_\mu\,u^*_\nu)\,F^{\mu\nu}
\nonumber \\&&
             +\frac{i\,e}{4}\,\biggl(Q+\frac{g-1}{m^2}\biggr)(u^*_{\mu\nu}\,u_\lambda - u_{\mu\nu}\,u^*_\lambda)
              \,\partial^\lambda F^{\mu\nu}
\label{02}
\end{eqnarray}
where
$u_{\mu\nu} = \nabla_\mu u_\nu - \nabla_\nu u_\mu$,
$\nabla_\mu = \partial_\mu + i\,e\,A_\mu$, and
$F^{\mu\nu} = \partial^\mu A^\nu - \partial^\nu A^\mu$.
The effective nonrelativistic Hamiltonian obtained from Eq. (\ref{02})
gives the $\vec\nabla\cdot\vec E$ term with the coefficient $e\,Q/6$ \cite{zatorski}.
What part of this coefficient should be included in the charge radius,
and what in the kinematical term $\delta_I$ ? The most natural assumption is that
$\left\langle R^2\right\rangle=0$ for a point $I=1$ particle.
One possible choice for a point vector particle
is $g=1$, $Q=0$, and another one $g=2$, $Q = -1/m^2$. 
The second choice
has the advantage that it leads to the  renormalized QED theory
of a charged vector particle,  while the first choice
leads to the simplest form of the Lagrangian. In this work,
following \cite{ch_radius}, we adopt the first choice, 
and consequently assume for a vector particle
$\delta_I = 0$. This assumption affects the relativistic recoil correction, 
while the finite nuclear size correction is
\begin{equation}
E_{FS} = \frac{2\,\pi\,\alpha}{3}\,\phi^2(0)\langle R^2\rangle,
\end{equation}
where $\phi^2(0) = (m_r\,\alpha)^3/(\pi\,n^3)\,\delta_{l0}$,
and $m_r$ is the $\mu$D reduced mass.

Having defined the leading finite size effect, we proceed
to the evaluation of nuclear structure corrections.
A nucleus is not a rigid particle, it can be excited by an orbiting
electron or muon, which results in the shift of atomic energies.
In the following, we derive general formulas for the nuclear 
polarizability shift with any Hamiltonian for deuterium, 
using the perturbation expansion
in the muon mass $m$ over the deuteron mass $m_D$,
including the so called Coulomb correction. 
Here, the deuteron binding energy counts as $m^2/m_D$.
With the assumed accuracy relativistic effects for the muon, 
as well as for the deuteron,
can be treated perturbatively and we start derivation from  the leading
electric dipole excitations. The nonrelativistic formula for the electric dipole
nuclear (scalar) polarizability correction is
\begin{equation}
\delta E = \alpha^2\,\biggl\langle\!\phi\,\phi_D\biggl|
\frac{\vec d\cdot\vec r}{r^3}\,
\frac{1}{E_D + E_0 - H_D -H_0}\,
\frac{\vec d\cdot\vec r}{r^3}\,
\biggr|\phi\,\phi_D\!\biggr\rangle 
\end{equation}
where $H_0$ is the nonrelativistic Coulomb
Hamiltonian for the muon with reduced mass $m_r$. 
Denoting the nuclear excitation energy by $E$,
the polarizability correction is
\begin{equation}
\delta E = \frac{\alpha^2}{3}\,\int_{E_T}dE\,
|\langle\phi_D|\vec d|E\rangle|^2
\biggl\langle\phi\biggl|
\frac{\vec r}{r^3}\,\frac{1}{E_0-H_0-E}\,
\frac{\vec r}{r^3}\,\biggr|\phi\biggr\rangle
\end{equation}
The nuclear excitation energy $E$ is much larger than a
typical atomic (muonic deuterium)  excitation energy,
thus one may perform the large $E$ expansion of the
electronic matrix element. The appropriate formula
for this expansion in atomic units, ${\cal E}=E/(m_r\,\alpha^2)$ is
\begin{align}
&\biggl\langle\phi\biggl|\frac{\vec r}{r^3}\,\frac{1}{H_0-E_0+{\cal E}}\,
\frac{\vec r}{r^3}\,\biggr|\phi\biggr\rangle \nonumber \\ &
= 4\,\pi\,\phi^2(0)\sqrt{\frac{2}{\cal E}} 
  + \frac{c_1}{\cal E} 
  - \frac{c_2}{\cal E}\,\sqrt{\frac{2}{\cal E}}
  + O({\cal E}^{-2}),
\end{align}
where 
\begin{align}
c_1(2S) &= \frac{1}{8} + \frac{1}{2}\,\ln\biggl(\frac{2}{\cal E}\biggr),&\;
c_2(2S) &= \frac{21}{32}+\frac{\pi^2}{12}, \nonumber \\
c_1(2P) &= \frac{1}{24},& c_2(2P) &= \frac{1}{16}.
\end{align}
From this expansion,
the leading electric dipole polarizability correction is \cite{friar2}
\begin{equation}
\delta_0 E =
-\frac{4\,\pi\,\alpha^2}{3}\,\phi^2(0)\int_{E_T}dE\,
\sqrt{\frac{2\,m_r}{E}}\,|\langle\phi_D|\vec d|E\rangle|^2,
\label{08}
\end{equation}
while the leading Coulomb correction, written explicitly for $2P-2S$ transition, 
is
\begin{eqnarray}
\delta_{C1} E &=& \frac{\alpha^6\,m_r^4}{6}\!
\int_{E_T}\frac{dE}{E}\,|\langle\phi_D|\vec d|E\rangle|^2\,
\biggl[\frac{1}{6}+\ln\biggl(\frac{2\,m_r\,\alpha^2}{E}\biggr)\biggr]
\nonumber \\ &=&
\frac{\alpha^5\,m_r^4}{4}\,\alpha_E\,\biggl[\frac{1}{6}
+\ln\biggl(\frac{2\,m_r\,\alpha^2}{\bar E}\biggr)\biggr]
\end{eqnarray}
where $\alpha_E$ is the electric dipole polarizability of a nucleus
\begin{equation}
\alpha_E = \frac{2\,\alpha}{3}\,\int_{E_T}\frac{dE}{E}\,
|\langle\phi_D|\vec d|E\rangle|^2 
\end{equation}
and $\bar E$ is the mean excitation energy. Both of them
have already been accurately calculated for the deuteron, namely 
$\alpha_E = 0.6330(13)$ fm$^3$ and $\bar E = 4.94$ MeV \cite{friar4}.
Our numerical results obtained in this work are in an agreement
with them. We shall mention that the formula for the leading Coulomb 
correction in $1S$ and $2S$ states was first obtained by Friar in \cite{friar2}. 

The next to leading Coulomb correction, which has not been considered so far,
written explicitly for $2P-2S$ transition, 
is
\begin{equation}
\delta_{C2} E =-\frac{1}{6}\,\biggl(\frac{19}{32}+\frac{\pi^2}{12}\biggr)\,
\alpha^7m_r^3\!\!\int_{E_T} \!\!\!dE\biggl(\frac{2\,m_r}{E}\biggr)^{3/2}\,
\!\!|\langle\phi_D|\vec d|E\rangle|^2 
\end{equation}
 The dipole operator $\vec d$ is the position
$\vec R$ of the proton with respect to the deuteron mass center.
It is thus assumed that there are no corrections
to the electric dipole operator, and for example mass center of the proton 
coincides with the charge center of the proton within the nucleus.
The uncertainty introduced by this approximation is unknown.
This is due to the fact that the underlying QCD theory
is nonperturbative and the resulting exact theory of nuclear forces
is not yet known. 

In the evaluation of further corrections we take the infinite nuclear mass limit,
thus neglecting nuclear recoil.
The relativistic corrections to the electric dipole polarizability effects
can be obtained from the two-photon exchange amplitude \cite{erickson}
\begin{eqnarray}
\delta E &=&
i\,e^4\,\phi^2(0)\,\frac{1}{3}\,\int_{E_T} dE \langle\phi_D|\vec d|E\rangle^2\,
\int\frac{d\,\omega}{2\,\pi}\,
\nonumber \\ &&
\int\frac{d^3k}{(2\,\pi)^3}\,\frac{1}{E+\omega}
\biggl(1+\frac{2\,\omega^4}{(\omega^2-k^2)^2}\biggr)\nonumber \\ && \times
\frac{4}{(\omega^2+2\,m\,\omega-k^2)(\omega^2-2\,m\,\omega-k^2)}.
\end{eqnarray}
The nuclear excitation energy $E$ is much smaller than the muon mass $m$ and
the nonrelativistic contribution comes from the region $\omega\approx E$ and
$k\approx\sqrt{2\,m\,E}$. Thus to obtain it one neglects $\omega^2$ and $\omega^4$.
The leading nonrelativistic term agrees with that in Eq. (\ref{08}), with the reduced mass $m_r$
replaced by a muon mass $m$. Relativistic correction comes from the next terms 
in the small $E$ expansion, namely
\begin{eqnarray}
\delta_R E &=& i\,e^4\,\phi^2(0)\,\frac{1}{3}\,\int_{E_T} dE \langle\phi_D|\vec d|E\rangle^2\,
\int\frac{d\,\omega}{2\,\pi} \nonumber \\ &&
\int\frac{d^3k}{(2\,\pi)^3}\,\frac{1}{E+\omega}
\frac{8\,\omega^2\,k^2}{(k^2-2\,m\,\omega)^2(k^2+2\,m\,\omega)^2}\nonumber \\
           &=& \frac{2\,\pi\,\alpha^2}{3}\,\phi^2(0)
\int_{E_T}\!\!dE\,\sqrt{\frac{E}{2\,m}}\,|\langle\phi_D|\vec d|E\rangle|^2.
\end{eqnarray}
This is the only relativistic correction which is not negligible
at our level of accuracy.

The corrections due to higher multipole polarizabilities and
higher order corrections due to the finite deuteron size can be treated together,
and we show that they cancel each other at the leading order.
Let us consider at first the related electronic matrix element $P$
for the nonrelativistic two-photon exchange
\begin{equation}
P =\biggl\langle\phi\biggl|\frac{\alpha}{|\vec r-\vec R|}
\frac{1}{(H_0-E_0+E)} \frac{\alpha}{|\vec r-\vec R'|}\biggr|\phi\biggr\rangle,
\end{equation}
where $H_0$ is the nonrelativistic Hamiltonian for the muon
in the nonrecoil limit, and $\vec R$ is a position of proton with respect
to the nuclear mass center. Using on  mass shell approximation
with subtractions of  the leading Coulomb interaction, the finite size,
and the electric dipole contribution, it becomes
\begin{eqnarray}
P &=& \alpha^2\,\phi^2(0)\,
\int\frac{d^3 q}{(2\,\pi)^3}\,\biggl(\frac{4\,\pi}{q^2}\biggr)^2\,
\biggl(E+\frac{q^2}{2\,m}\biggr)^{-1}\nonumber \\ &&
\biggl[e^{i\,\vec q\cdot(\vec R-\vec R')}-1+\frac{q^2}{6}\,(\vec R-\vec R')^2\biggr]
\label{15}\\&\approx&
\frac{\pi}{3}\,m\,\alpha^2\,\phi^2(0)\,|\vec R-\vec R'|^3
\biggl(1-\frac{\sqrt{2\,m\,E}}{5}\, |\vec R-\vec R'|\biggr), \nonumber
\end{eqnarray}
where we performed expansion in the small parameter $\sqrt{2\,m\,E} |\vec R-\vec R'|$.
The corresponding correction to atomic energy is
\begin{equation}
\delta_Q E = -\sum\hspace{-3.0ex}\int dE\!\!\int\!\!d^3 R\,d^3R'\,
\phi^*_D(\vec R)\,\phi_E(\vec R)\,\phi_D(\vec R')\,\phi^*_E(\vec R)\,P
\end{equation}
Consider now the first $E$-independent term. When $\phi_E = \phi_D$,
it will be the elastic part know as a Zemach correction \cite{friar3},
but the inclusion of all excited states leads to $\delta(\vec R-\vec R')$
and the $|\vec R-\vec R'|^3$ term vanishes completely.
As a result, there is no Zemach [elastic $O(\alpha^5)$] 
correction for muonic deuterium. 
From Eq. (\ref{15}) only the second term  remains, which gives
\begin{eqnarray}
\delta_Q E &=& \frac{2\,\pi}{15}\,m^2\,\alpha^2\,\phi^2(0)
\!\int_{E_T} \!\!dE\,\sqrt{\frac{E}{2m}}\int \!\!d^3 R\,d^3R'
\\ &&
\phi^*_D(\vec R)\,\phi_E(\vec R)\,\phi_D(\vec R')\,
\phi^*_E(\vec R')\,[(\vec R-\vec R')^2]^2. \nonumber
\end{eqnarray}
These corrections are due to
the electric dipole, the quadrupole and the monopole nuclear excitations,
namely
\begin{eqnarray}
\delta_Q E &=& \frac{2\,\pi}{15}\,m^2\,\alpha^2\,\phi^2(0)
\!\int_{E_T} \!\!dE\,\sqrt{\frac{E}{2m}}
\nonumber \\ &&
\biggl[\frac{10}{3}\,\langle\phi_D|R^2|E\rangle^2
-8\,\langle\phi_D|R^i|E\rangle\,\langle E|R^2\,R^i|\phi_D\rangle
\nonumber \\ &&
+4\,\langle\phi_D|(R^i\,R^j-\delta^{ij}\,R^2/3)|E\rangle^2\biggr]
\\ &=& \delta_{Q0} E + \delta_{Q1} E + \delta_{Q2} E. \nonumber
\end{eqnarray}

As we assumed at the beginning, all corrections
of order $\alpha^5\,m^2/m_D$ are neglected.
However, due to the large magnetic moment anomaly
of the proton and the neutron, we make an exception and consider the magnetic dipole
polarizability correction. It comes from
\begin{eqnarray}
H_{M1} &=& -\vec\mu\cdot\vec B = -\frac{e}{2\,m_p}(g_p\,\vec s_p+g_n\,\vec s_n)\cdot \vec B
\nonumber \\ &\approx&
-\frac{e\,(g_p-g_n)}{2\,m_p}\,\frac{(\vec s_p-\vec s_n)}{2}\cdot\vec B,
\end{eqnarray}
where $g_p = 5.586$, and $g_n = -3.826$.
We use the analogous two-photon exchange formula as for the electric dipole transitions
%\begin{widetext}
\begin{eqnarray}
\delta_M E &=&
i\,e^2\,\phi^2(0)\,
\int\frac{d\,\omega}{2\,\pi}\,\int\frac{d^3k}{(2\,\pi)^3}\,
\frac{\delta^{ik}}{\omega^2-k^2}\,
\frac{\delta^{jl}}{\omega^2-k^2}
\nonumber \\ && \hspace{-7ex}\times
{\rm Tr}\biggl[\biggl(\gamma^j\,\frac{1}{\not\!p-\not\!k-m}\gamma^i +
\gamma^i\frac{1}{\not\!p+\not\!k-m}\,\gamma^j\biggr)\,\frac{(\gamma^0+I)}{4}\biggr]\,
\nonumber \\ && \hspace{-7ex}\times
\biggl\langle\phi_D\biggl|(\vec\mu\times\vec k)^{\,k}\,\frac{1}{E_D-H_D-\omega}\,
(\vec\mu\times\vec k)^{\,l}\biggr|\phi_D\biggr\rangle
\end{eqnarray}
%\end{widetext}
and perform the nonrelativistic approximation:
\begin{eqnarray}
\delta_M E &=&
\frac{8\,\,\pi\,\alpha^2}{3}\,\phi^2(0)
\biggl(\frac{g_p-g_n}{4\,m_p}\biggr)^2
\nonumber \\ && \times
\int_{E_T}\!\!\!\!dE\,\sqrt{\frac{E}{2\,m}}\,
\langle\phi_D|\vec s_p-\vec s_n|E\rangle^2
\end{eqnarray}
Corrections due to the intrinsic proton polarizability and the proton Zemach moment
are to a good approximation the same as in muonic hydrogen. Therefore,
we use a recent result of very thorough calculations \cite{carlson} 
$\Delta E(2S) = -36.9(2.4)\,\mu$eV and scale it by a factor 
\begin{equation}
\delta_P E = - \Delta E(2S)\,m_{rD}^3/m_{rH}^3.
\end{equation}
The final expression for the nuclear polarizability combined with
the elastic contribution is
\begin{equation}
\Delta E = \delta_0 E + \delta_C E + \delta_R E + \delta_Q E + \delta_M E+ \delta_P E.
\end{equation}

Numerical calculation of deuteron matrix elements are performed
using the discrete variable representation (DVR) \cite{dvr} method.
In the DVR method, the Hamiltonian is represented as a symmetric matrix, which 
can be diagonalized and all formulas represented as a finite sum
over the spectrum. Numerical results using the modern  nucleon-nucleon 
AV18 potential from Argonne National Laboratory \cite{av18},
are presented in Table I.
\begin{table}[!htb]
\caption{Nuclear structure corrections in muonic deuterium
for $2P-2S$ transition.}
\label{table1}
\begin{ruledtabular}\as{1.25}
 \begin{tabular}{cw{2.6}}
\centt{correction}&   \centt{value in meV}\\ 
\hline
$\delta_0 E$      &  1.910    \\
$\delta_{C1} E$   & -0.255   \\
$\delta_{C2} E$    & -0.006   \\
$\delta_R E$      & -0.035    \\
$\delta_{Q0} E$    & -0.045  \\
$\delta_{Q1} E$    &  0.151   \\
$\delta_{Q2} E$    & -0.066  \\
$\delta_M E$      &  -0.016  \\
$\delta_P E$      &  0.043(3)  \\[1ex]
$\Delta E $       & 1.680(16)
\end{tabular}
 \end{ruledtabular}
\end{table}
We have checked numerics by the calculation of the electric dipole polarizability, 
and out result $\alpha_d= 0.634$ fm$^3$ is  close to the recommended value 
from \cite{friar4}. The difference of about $0.001$ fm$^3$ comes from the fact
that we take into account a small neutron-proton mass difference in the electric 
dipole operator. 

Surprisingly, the total value $\Delta E $  is close to the
nonrelativistic electric dipole polarizability contribution
$\delta_0 E + \delta_{C1} E + \delta_{C2} E = 1.649$ meV.
This means that relativistic and higher multipole corrections,
although individually not small, tend to cancel between themselves.
For the final uncertainty, we assume 50\% of these higher order 
corrections. We can not at this moment give a more reliable 
estimate of uncertainty, but note that it is about 20 times smaller 
than the discrepancy in muonic hydrogen.

Our result for the nuclear structure correction $\Delta E$ 
is not in good agreement with former calculations.
Leidemann and Rosenfelder in \cite{rosenfelder} obtained 
for the polarizability correction of 2S state the result $-1.500(25)$ meV. 
This should be combined with the elastic contribution obtained  by Martynenko 
\cite{martynenko1} of $-0.37$ meV and the Coulomb correction of $0.26$ meV, 
which totals to $-1.61(3)$ meV.
This is $2\sigma$ away from our result, shown in Table I. 
The difference may come from three sources.
The first one is lack of clear separation between the elastic contribution 
from \cite{martynenko1} and the inelastic one from \cite{rosenfelder}, 
thus some terms might be counted twice. The second one is neglect in \cite{rosenfelder}
of the intrinsic proton polarizability correction of $0.013$ meV \cite{carlson}.
The third source is the extra coefficient $R_\mu=0.9778$ used in \cite{rosenfelder}
for the polarizability correction, which reflects the fact that probability of 
finding muon within the deuteron is not exactly $\phi^2(0)$ but $R_\mu\,\phi^2(0)$. 
To verify this coefficient one has to investigate  
three photon exchange correction, 
%(apart that of $\delta_{C1}E$),
details are beyond the scope of this work
but we claim lack of such coefficient.

In order to shed light on the proton charge radius discrepancy,
we consider the difference
%\begin{equation}
$E_D(2P-2S) - E_H(2P-2S)\,m_{rD}^3/m_{rH}^3, $
%\label{24}
%\end{equation}
where the proton size and the proton polarizability cancel out.
This difference is sensitive to the deuteron structure radius, 
which is known from a very accurate H-D$(2S-1S)$ isotope shift.
If agreement between experiment and theoretical predictions
based on QED calculations \cite{borie, martynenko2}
including nuclear polarizability correction calculated in this work
is observed, this may mean that the muonic hydrogen discrepancy
is caused by a local ($\sim$ fm) muon-proton interaction
or by a 5$\sigma$ shift in the Rydberg constants \cite{pohl}.
There could be as well a different source of the discrepancy,
a long-range type interaction ($\sim 200$ fm), and this will not
cancel out in the difference (\ref{24}), and a small
discrepancy in $\mu$D will persist in this case.

In summary, we have demonstrated in this work 
that accurate predictions for muonic deuterium are feasible,
the non-QED corrections have been here accurately calculated,
and comparison with experimental transitions in $\mu$D
will give hints on a possible source of discrepancies 
in the proton charge radius. 

\begin{acknowledgments}
KP acknowledges support by NIST through Precision Measurement Grant
PMG 60NANB7D6153.
\end{acknowledgments}


\begin{thebibliography}{99}
\bibitem{NIST} P. J. Mohr, B. N. Taylor, and D. B. Newell, Rev. Mod. Phys. {\bf 80}, 633 (2008).
\bibitem{rp_mainz} J. C. Bernauer {\em et al.}, Phys. Rev. Lett. {\bf 105}, 242001 (2010).
\bibitem{rp_jlab} X. Zhan {\em et al}., (Jefferson Lab Hall A Collaboration), arxiv:1102.0318 [nucl-ex]. 
\bibitem{H-Da} A. Huber {\em et al.}, Phys. Rev. Lett. {\bf 80}, 468 (1998). 
\bibitem{H-Db} C. G. Parthey {\em et al}., Phys. Rev. Lett. {\bf 104}, 233001 (2010).
\bibitem{be11} W. N\"ortersh\"auser {\em et al.}, Phys. Rev. Lett. {\bf 102}, 062503 (2009). 
\bibitem{pohl} R. Pohl {\em et al.}, Nature {\bf 466}, 213 (2010).
\bibitem{spin1} J.A. Young and S.A. Bludman, Phys. Rev. {\bf 131}, 2326 (1963).
\bibitem{zatorski} J. Zatorski and K. Pachucki, Phys. Rev. A {\bf 82}, 052520 (2010).
\bibitem{ch_radius} J.L. Friar, J. Martorell, and D.W.L. Sprung, Phys. Rev. A {\bf 56}, 4579 (1997).
\bibitem{friar4}J. L. Friar, G. L. Payne, Phys. Rev. C {\bf 55}, 2764 (1997);
                                          Phys. Rev. C {\bf 72} 014004 (2005). %deuteron polarizability
\bibitem{friar2} J.L. Friar, Phys. Rev. C {\bf 16}, 1540 (1977). %muHe
\bibitem{erickson} J. Bernab\'eu and T.E.O. Ericson, Z. Phys. A {\bf 309}, 213 (1983).
\bibitem{friar3} J. L. Friar, Ann. Phys. 122, 151 (1979). %Zemach moment
\bibitem{carlson} C. E. Carlson and M. Vanderhaeghen, arxiv:1101.5965 [hep-ph].
\bibitem{dvr} D. T. Colbert and W. H. Miller, J. Chem. Phys. {\bf 96}, 1982 (1992).
\bibitem{av18} R.B. Wiringa, V.G.J. Stoks, and R. Schiavilla, Phys. Rev. C {\bf 51}, 38 (1995).
%\bibitem{rosenfelder} R. Rosenfelder, Phys.Lett. {\bf B463}, 317 (1999). 
\bibitem{rosenfelder} W. Leidemann and R. Rosenfelder, Phys. Rev. C {\bf 51}, 427 (1995).
\bibitem{martynenko1} A.P. Martynenko and R.N. Faustov, Phys. Atom. Nucl. {\bf 67}, 457 (2004), 
                      [Yad. Fiz. {\bf 67}, 477 (2004)]. %nuclear structure in muD
\bibitem{borie} E. Borie, Phys. Rev. A {\bf 72}, 052511 (2005).
\bibitem{martynenko2} A.P. Martynenko, J. Exp. Theor. Phys. {\bf 101}, 1021 (2005).
\end{thebibliography}
\end{document}